\documentclass[aps,prx,superscriptaddress,twocolumn,longbibliography]{revtex4-1}
\usepackage{bm} \usepackage{graphicx} \usepackage{amsmath}
\usepackage{bbm}
\usepackage{amsthm,stmaryrd}
\usepackage{amssymb} \usepackage{epstopdf} 
\usepackage{color}
\usepackage{amsmath}
\usepackage{hyperref}
\usepackage{comment}
\newcommand{\ket}[1]{|{#1}\rangle}
\newcommand{\bra}[1]{\langle{#1}|}

\graphicspath{ {./figures/} }

\DeclareMathOperator*{\Tr}{Tr}

\DeclareMathOperator{\diag}{diag}
\DeclareMathOperator{\id}{\mathbbm{1}}
\DeclareRobustCommand\openzero{\leavevmode\hbox{0\kern-.55em0}}
\mathchardef\minus="002D

\definecolor{JM}{RGB}{4,116,149}

\newcommand{\rvline}{\hspace*{-\arraycolsep}\vline\hspace*{-\arraycolsep}}

\begin{document}

\newtheorem*{prop}{Proposition}

\title{
	Training Gaussian Boson Sampling Distributions
}

\author{ Leonardo Banchi}
\affiliation{ Department of Physics and Astronomy, University of Florence, via G. Sansone 1, I-50019 Sesto Fiorentino (FI), Italy}
\affiliation{ INFN Sezione di Firenze, via G.Sansone 1, I-50019 Sesto Fiorentino (FI), Italy }

\author{ Nicol\'as Quesada }
\affiliation{Xanadu, Toronto, ON, M5G 2C8, Canada
	}

\author{Juan Miguel Arrazola}
\affiliation{Xanadu, Toronto, ON, M5G 2C8, Canada
	}

\date{\today}
\begin{abstract}
Gaussian Boson Sampling (GBS) is a near-term platform for photonic quantum
computing. Applications have been developed which rely on directly programming GBS devices, but the ability to train and optimize circuits has been a key missing ingredient for developing new algorithms. In this work, we derive analytical gradient formulas for the GBS distribution, which can be used to train devices using standard methods based on gradient descent. We introduce a parametrization of the distribution that allows the gradient to be estimated by sampling from the same device that is being optimized. In the case of training using a Kullback-Leibler divergence or log-likelihood cost function, we show that gradients can be computed classically, leading to fast training. We illustrate these results with numerical experiments in stochastic optimization and unsupervised learning.  As a particular example, 
we introduce the variational Ising solver, a hybrid algorithm for training GBS devices to sample ground states of a classical Ising model with high probability. 
\end{abstract}

\maketitle

\section{Introduction}
Gaussian Boson Sampling (GBS) is a special-purpose platform for photonic quantum computing. It was proposed as a method to build photonic devices capable of performing tasks that are intractable for classical computers \cite{hamilton2017gaussian, kruse2019detailed}. Since then, several quantum algorithms based on GBS have been introduced~\cite{bromley2019applications}, with applications to graph optimization~\cite{arrazola2018using, arrazola2018quantum, banchi2019molecular}, graph similarity \cite{bradler2018graph, schuld2019quantum}, point processes \cite{jahangiri2019point}, and quantum chemistry \cite{huh2015boson,huh2017vibronic}. These algorithms rely on strategies to carefully program GBS devices, typically by encoding a suitable symmetric matrix into the GBS distribution.

Yet many quantum algorithms rely on the ability to train the parameters of quantum circuits~\cite{mcclean2016theory}, a strategy inspired by the success of neural networks in machine learning. Examples include quantum approximate optimization \cite{farhi2014quantum, zhou2018quantum}, variational quantum eigensolvers \cite{peruzzo2014variational}, quantum feature embeddings \cite{schuld2019feature, havlivcek2019supervised}, and quantum classifiers~\cite{schuld2018circuit}. Training is often performed by evaluating gradients of a cost function with respect to circuit parameters, then employing gradient-based optimization methods~\cite{bergholm2018pennylane, schuld2019evaluating}. Deriving similar methods to train GBS devices is a missing piece for unlocking new algorithms, particularly in machine learning and optimization.

In this work, we derive analytic gradients of the GBS distribution which can be used to train the device using gradient-based optimization. We derive a general gradient formula that can be evaluated in simulators, but is not always accessible from hardware. We then introduce a specific parametrization of the GBS distribution that expresses the gradient as an expectation value from the same distribution. Such gradients can be evaluated by sampling from the same device that is being optimized. Using this parametrization, we show that for Kullback-Leibler divergence or log-likelihood cost functions, analytical gradients can be evaluated efficiently using classical methods, leading to fast training. We illustrate these results with numerical experiments in stochastic optimization and unsupervised learning.

As a specific application for our training scheme, we introduce the variational Ising solver (VIS). In this algorithm, as in the variational quantum eigensolver \cite{peruzzo2014variational}, a parametric circuit is optimized to approximate the ground state of a Hamiltonian. Similarly to the quantum approximate optimization algorithm \cite{farhi2014quantum,zhou2018quantum,gentini2019noise}, we focus on combinatorial optimization problems where the Hamiltonian can be expressed as a classical Ising model. Both the variational eigensolver and the quantum approximate optimization algorithm are tailored for near-term 
qubit-based quantum computers, while VIS is tailored for near-term GBS devices. We use a parametric circuit that creates a particular Gaussian state, and iteratively update the Gaussian state 
using a gradient-based hybrid strategy based on outcomes coming from either photon-number-resolving detectors or threshold detectors. 

The paper is organized as follows. In Sec.~\ref{Sec:GBS}, we provide a short review of GBS. In Sec.~\ref{Sec:Training}, we discuss mathematical details of the stochastic optimization and unsupervised learning tasks covered in this work. Sec.~\ref{Sec:Gradients} presents the analytical gradient formulas and parametrizations of the GBS distribution, as well as some of its extensions. Finally, in Sec.~\ref{Sec:Apps}, we provide numerical examples demonstrating the ability of VIS to approximate the solution to certain combinatorial optimization problems, and the ability 
to train GBS distributions using classical gradient formulas.
Conclusions are drawn in Sec.~\ref{Sec:Conclusions}.

\section{Gaussian Boson Sampling}\label{Sec:GBS}

In quantum optics, the systems of interest are optical modes of the
quantized electromagnetic field. The quantum state of $m$ modes can be specified by its Wigner function $W(\bm{q}, \bm{p})$, where $\bm{q},\bm{p}\in \mathbb{R}^m$ are known respectively as the position
and momentum quadrature vectors. Gaussian states are characterized by having a Wigner function that is Gaussian. Consequently, Gaussian states can be completely specified by their first and second 
moments, namely two $m$-dimensional vectors of means $\bm{\bar{q}}, \bm{\bar{p}}$ and a covariance matrix $\Sigma$. For our purposes, it is more convenient to work with the complex-normal random variable $\bm{\alpha} = \tfrac{1}{\sqrt{2 \hbar}} (\bm{q}+i \ \bm{p})$ that has mean $\bm{\bar{\alpha}} = \tfrac{1}{\sqrt{2 \hbar}} (\bm{\bar{q}}+i \ \bm{\bar{p}})$ and covariance matrix $V$.

When measuring a Gaussian state in the photon-number basis, the probability of observing an outcome $\ket{\bar n}=\ket{n_1,\dots,n_m}$, where $n_i$ is the number of photons in mode $i$, is given by 
 \cite{hamilton2017gaussian}:
\begin{equation}
	P_\mathcal{A} 
	(\bar n) = 
	\frac{1}{\mathcal{Z}}\frac{{\rm Haf} (\mathcal 
	A_{\bar n \oplus \bar n})}{n_1!\cdots n_m!},
	\label{e:GBSx}
\end{equation}
where 
\begin{align}
\mathcal A &= X \left(\id - (V+\id/2)^{-1}\right),
\label{e:amat} \\
X &:=  \left[\begin{smallmatrix}
	0 &  \id \\
	\id & 0  
\end{smallmatrix} \right],\\
\frac{1}{\mathcal{Z}}&:= \sqrt{\det(\id  - X\mathcal A)}.
\end{align}
For a matrix $\mathcal{B} \in \mathbb{C}^{m \times m}$ and outcome vector $\bar{n} = (n_1,\ldots,n_m)$, the notation $\mathcal{B}_{\bar{n}}$
indicates the matrix constructed from $\mathcal{B}$ as follows. If $n_i = 0$, the $i^{\text{th}}$ row and column are deleted from $\mathcal{B}$. If $n_i > 0$, the $i^{\text{th}}$ row and column are repeated $n_i$ times. In the case of $\mathcal{A} \in \mathbb{C}^{2m \times 2m}$ as in Eq.~\eqref{e:GBSx}, the outcome vector is $\bar{n} \oplus \bar{n} = (n_1,\ldots,n_m,n_1,\ldots,n_m)$.

The hafnian of a $2m\times 2m$ matrix $\mathcal{A}$ is defined as \cite{caianiello1953quantum}
\begin{equation}
{\rm Haf}(\mathcal{A}) = \sum_{\mu\in {\rm PMP}(2m)} \prod_{(i,j)\in \mu} \mathcal{A}_{i,j},
\end{equation}
where $\mathcal{A}_{i, j}$ is the $(i,j)$ entry of the symmetric matrix $\mathcal{A} = \mathcal{A}^T$ and $\rm PMP$ is the set of perfect matching permutations, the possible ways of partitioning the set $\{1,\dots,2m\}$ into disjoints subsets of size two. The hafnian is \#P-Hard to approximate for worst-case instances  \cite{barvinok2016combinatorics} and the runtime of the best known algorithms for computing hafnians of arbitrary matrices scales exponentially with $m$ \cite{bjorklund2018faster}. Using techniques from Ref.~\cite{aaronson2013},  it has been argued that sampling from a GBS distribution cannot be done in classical polynomial time unless the polynomial hierarchy collapses to third level~\cite{hamilton2017gaussian}.

For pure Gaussian states, it holds that $\mathcal A = A\oplus A^*$ and $A \in \mathbb{C}^{m \times m}$ is a symmetric matrix that can be decomposed as
\begin{equation}
A = U \diag(\lambda_1, \ldots, \lambda_m) U^T,
\end{equation}
where $0\leq \lambda_i<1$. The probability distribution is then 
\begin{equation}
	P_\mathcal{A}(\bar n) = 
	\frac{1}{\mathcal{Z}}\frac{|{\rm Haf} (A_{\bar n})|^2}{n_1!\cdots n_m!}.
	\label{e:GBS0}
\end{equation}
The mean photon number is given by 
\begin{equation}
\langle n \rangle=\sum_{i=1}^m\frac{\lambda_i^2}{1-\lambda_i^2},
\end{equation}
which can be adjusted by rescaling the matrix $A\rightarrow cA$ for an appropriate parameter $c>0$.
\section{Training the GBS distribution}\label{Sec:Training}
In this section, we describe the training tasks considered in this work: stochastic optimization and unsupervised learning. Here and throughout the manuscript, given a vector of parameters $\theta=(\theta_1,\theta_2,\ldots,\theta_d)$, we use $\partial_\theta$ as a shorthand for the gradient $(\frac{\partial}{\partial\theta_1},\frac{\partial}{\partial\theta_2},\ldots,\frac{\partial}{\partial\theta_d})$. Similarly, we employ $\partial_{\theta_j}$ to denote $\frac{\partial}{\partial\theta_j}$.

\subsection{Stochastic optimization}
It has been recently shown that certain optimization problems in graph theory can be solved by sampling solutions from 
a properly configured GBS device~\cite{arrazola2018using,banchi2019molecular}. 
This was made possible by encoding graphs into the GBS distribution
\cite{bradler2018gaussian} and exploiting the fact that this distribution 
outputs, with high probability, photon configurations $\bar{n}$ that have a large hafnian ${\rm Haf}(\mathcal A_{\bar n})$. 

We consider the more general problem of optimizing the GBS distribution 
directly from the samples, without requiring a theoretical scheme to optimally 
program the device.  Consider 
a function $H(\bar n)$ that associates a cost to the set 
of positive integers $n_k$ sampled from the GBS distribution. 
Fixing the symmetric matrix $\mathcal A = \mathcal A(\theta)$ 
where $\theta$ is a set of variational parameters, the cost is given by 
\begin{equation}
	C(\theta) = \mathbb E_{\bar n\sim  P_{\mathcal{A}(\theta)}(\bar n)}[H(\bar n,\theta)] \equiv 
	\sum_{\bar n} H(\bar n) P_{\mathcal{A}(\theta)}(\bar n)~.
	\label{e:cost}
\end{equation}
Our goal is to optimize the Gaussian state, encapsulated by the $2M\times 2M$ matrix $\mathcal A(\theta)$, in order to minimize the cost function. 
Suppose that there are certain choices of the parametrization such that the gradient 
 $\partial_{\theta} C(\theta)$ can be either efficiently computed numerically or 
 estimated via sampling on a physical device. In such cases it is possible to 
 minimize the average cost $C(\theta)$ using the update rule
\begin{equation}\label{e:update}
\theta \rightarrow \theta - \eta\, \partial_{\theta} C(\theta),
\end{equation}
where $\eta>0$ is a learning rate. Alternatively, other gradient-based optimization algorithms can be used \cite{bubeck2015convex,spall2005introduction}.

We show that, for some parametrizations of the Gaussian state, it is possible to write 
\begin{equation}
		\partial_\theta C(\theta) = 
		\mathbb E_{\bar n\sim  P_{\mathcal{A}'(\theta)}(\bar n)}[G(\bar n)], 
	\label{e:costgbsgrad}
\end{equation}
namely it is possible to write the gradient of $C(\theta)$ as an expectation value of 
a different function $G(\bar n)$ with respect to a possibly different GBS distribution 
$P_{\mathcal{A}'(\theta)}(\bar n)$. A GBS device can then be used to sample from this new distribution and 
obtain an empirical gradient 
\begin{equation}
	\partial_\theta C(\theta) \approx \frac 1T\sum_{t=1}^T G(\bar n^{(t)}),
	\label{e:empiricalgrad}
\end{equation}
from the samples $\{\bar{n}^{(1)}, \ldots, \bar{n}^{(T)} \}$. The parameters are then iteratively updated using the gradient estimate
\begin{equation}
	\theta \rightarrow \theta - \eta\, \frac 1T\sum_{t=1}^T G(\bar n^{(t)})~.
	\label{e:SGD}
\end{equation}

\subsection{Unsupervised learning}
In a standard unsupervised learning scenario, data are assumed to be sampled from 
an unknown probability distribution $Q(\bar n)$, and a common goal is to learn that 
distribution. This is done by considering a convenient model and updating its parameters such that the data sequence matches 
the samples from the model distribution $P(\bar n)$. Training can be performed by minimizing a suitably
chosen cost function, such as the Kullback-Leibler (KL) divergence
\begin{equation}
	D_{KL}[Q,P] = \sum_x Q(x)\log\frac{Q(x)}{P(x)}~.
	\label{e:kl}
\end{equation}

We study the KL divergence between a data distribution and a 
GBS distribution with parameters $\theta$:
\begin{equation}
	C_{\rm data}(\theta) = D_{KL}[P_{\rm data}(\bar n),P_{\mathcal{A}(\theta)}(\bar n)].
	\label{e:cdata}
\end{equation}
Its gradient is given by 
\begin{align}
	\partial_\theta C_{\rm data}(\theta) &= - \sum_{\bar n} P_{\rm data}(\bar n)\,
	\partial_{\theta} \log P_{\mathcal{A}(\theta)}(\bar n)
	\nonumber
	\\ &= 
	\mathbb{E}_{\bar n\sim P_{\rm data}}
\left[-
	\partial_{\theta} \log P_{\mathcal{A}(\theta)}(\bar n)
	\right]
	~.
	\label{e:cdata_grad}
\end{align}
In practice, instead of an explicit expression for the data distribution $P_{\rm data}(\bar n)$, a training set $\{\bar{n}^{(1)}, \ldots, \bar{n}^{(T)} \}$ is provided. This is interpreted as a collection of
samples from the data distribution. Averages are defined with respect to these samples:
\begin{equation}
\mathbb{E}_{\bar n\sim P_{\rm data}}
\left[-
	\partial_{\theta} \log P_{\mathcal{A}(\theta)}(\bar n)
	\right] = -\frac{1}{T}\sum_{t=1}^T\partial_{\theta} \log P_{\mathcal{A}(\theta)}(\bar n^{(t)}).
\end{equation}
We show that for certain choices of the parametrization, it is possible to 
compute the derivatives $\partial_{\theta} \log P_{\mathcal{A}(\theta)}(\bar n)$, 
allowing for an efficient training of the GBS distribution.

\section{Analytical gradients}\label{Sec:Gradients}
We describe gradient formulas for the GBS distribution. The first result is a general formula expressing the gradient for arbitrary parametrizations. We proceed by describing a strategy, the WAW parametrization, that allows gradients for arbitrary cost functions to be computed as expectation values over GBS distributions. Moreover, for specific cost functions, we show that gradients can be efficiently calculated classically. Finally, we discuss gradient formulas for GBS with threshold detectors, reparametrization strategies, and the projected subgradient method. 

\subsection{General formula}
The gradient of the GBS distribution in Eq.~\eqref{e:GBSx}, $P_{\mathcal A} (\bar n) = 
	\frac{1}{\mathcal{Z}}\frac{{\rm Haf} (\mathcal A_{\bar n \oplus \bar n})}{n_1!\cdots n_m!}$,
	can be expressed as
\begin{align}
\partial_\theta P_{\mathcal A}(\bar n) = \left(\partial_\theta \frac{1}{\mathcal{Z}}\right) \frac{{\rm Haf} (\mathcal 
	A_{\bar n \oplus \bar n})}{n_1!\cdots n_m!}+
	\frac{1}{\mathcal{Z}}\frac{\partial_\theta {\rm Haf} (\mathcal 
	A_{\bar n \oplus \bar n})}{n_1!\cdots n_m!}.
\label{e:gradP}
\end{align}
Note that in this section we avoid writing the explicit dependence of $\mathcal{A}$ on $\theta$ to simplify the notation. 
As shown in Appendix~\ref{a.haf}, the derivatives in Eq.~\eqref{e:gradP} can be calculated analytically and the result is
\begin{align}
	\partial_\theta \left(\frac{1}{\mathcal{Z}}\right)&=-\frac{1}{2}\Tr\left[\frac{1}{\mathcal{Z}}\frac{\partial_\theta \mathcal A}{X - \mathcal A}  \right],\label{e:gradZ}\\
\partial_\theta {\rm Haf} (\mathcal 
	A_{\bar n \oplus \bar n})&=\sum_{i\neq j}^{2N} 
	(\partial_\theta  \mathcal A_{\bar n})_{ij} \; {\rm Haf} \left( \mathcal 
A^{[i,j]}_{\bar n \oplus \bar n} \right),
\end{align}
where $2N$, with $N=\sum_k n_k$, is the dimension of the matrix $\mathcal A_{\bar n\oplus\bar n}$. The submatrix
 $A^{[i,j]}_{\bar n \oplus \bar n} $ is constructed from $A_{\bar n \oplus \bar n} $ by removing rows $(i,j)$ and columns $(i,j)$. 
Combining these results gives a general formula for the gradient of the GBS distribution:
\begin{align}
&\partial_\theta P_{\mathcal A}(\bar n)= 
 -  \frac{1}{2}\Tr\left[\frac{ \partial_\theta \mathcal A}{X - \mathcal A} \right]
P_{\mathcal A}(\bar n) + \nonumber
\label{e:diffGBS}
\\   &  \frac{1}{\mathcal{Z}}\frac{1}{n_1!\cdots n_m!} 
 \sum_{i\neq j}^{2N} 
 (\partial_\theta  \mathcal A_{\bar n\oplus \bar n})_{ij} \; {\rm Haf} \left( \mathcal A^{[i,j]}_{\bar n \oplus \bar n} \right).
\end{align}
From the above equation we can also obtain the derivative of the cost function $C(\theta)$ in Eq.~\eqref{e:cost}: 
\begin{align}
\partial_\theta C(\theta)&=\sum_{\bar{n}}H(\bar{n})\partial_\theta P_{\mathcal A}(\bar{n})\\
&=-\frac{1}{2}\mathbb{E}_{\bar n\sim P(\bar n)}\left[\Tr\left(\frac{H(\bar n)}{X - \mathcal A} \partial_\theta \mathcal A\right) \right]+\nonumber\\
&\frac{\mathcal{Z}^{-1}}{n_1!\cdots n_m!}
\sum_{\bar{n}}H(\bar n) \sum_{i\neq j}^{2N} 
(\partial_\theta  \mathcal A_{\bar n\oplus\bar n})_{ij} \; {\rm Haf} \left( \mathcal A^{[i,j]}_{\bar n \oplus \bar n} \right).
\end{align}
The generalization to a $\theta$-dependent cost function is straightforward. 

The quantities ${\rm Haf} \left( \mathcal A^{[i,j]}_{{\bar n \oplus \bar n }}\right)$ are not proportional to probabilities unless $i=j+N$ or $i+N=j$ \cite{quesada2019simulating}, which makes it challenging to express gradients as expectations over the GBS distribution. Nevertheless, as we describe next, it is possible to cast gradients as expectation values for carefully chosen parametrizations of the matrix $\mathcal A$.

\subsection{The WAW parametrization}
We focus on the pure-state case, $\mathcal A = A\oplus A^*$, and replace the matrix 
$A$ with 
\begin{align}
A_W &= W A W,\label{e:AW}
\end{align}
where $W_{kj} = \sqrt{w_k} \delta_{kj}$ and $w_k\geq0$. 
The generalization to mixed states is studied in Appendix~\ref{a.weight}. 
The symmetric matrix $A$ is kept fixed and the weights $w_k$ of the diagonal weight matrix $W$ are trainable parameters. The matrix $A$ serves as a model for the distribution and $W$ encodes its free parameters. We refer to this strategy as the WAW parametrization, in reference to Eq.~\eqref{e:AW}. Similar parametrizations have been succesfully used for training determinantal point processes in machine learning \cite{kulesza2011learning}.

It is important that when updating parameters, the matrix $A_W$ always corresponds to a physical 
Gaussian state. As shown in Appendix~\ref{a.weight}, if $A$ is a valid matrix with singular values contained in $[0,1)$, $A_W$ is also valid whenever $0 \leq w_k\leq 1$. This condition can be enforced via reparametrization. 
One of the strategies we consider is to express $w_k(\theta)$ as
\begin{equation}\label{e:reparam}
w_k(\theta) = \exp(-\theta^T f^{(k)}),
\end{equation}
where $f^{(k)}=(f^{(k)}_1, f^{(k)}_2, \ldots, f^{(k)}_d)$ is a $d$-dimensional vector, and $\theta=(\theta_1, \theta_2, \ldots, \theta_d)$ is a vector of parameters. The condition $0\leq w_k\leq 1$ can be satisfied by enforcing $\theta^T f^{(k)}\geq 0$ for all $k$.

The hafnian of $A_W$ can be factorized into independent contributions from $A$ and $W$~\cite{barvinok2016combinatorics}:
\begin{equation}
{\rm Haf}(A_W) = {\rm Haf}(A)\det(W).
\label{e:wid}
\end{equation}
Inserting the above in Eq.~\eqref{e:GBS0} gives 
\begin{align}
P_{A,W}(\bar n) &= 
\frac{1}{\mathcal{Z}} \; {\rm Haf} (A_{\bar n})^2 
\prod_{i=1}^{m}  \frac{w^{n_i }_{i}}{n_i!},\label{e:GBSW}
\end{align}
where the notation 
$P_{A,W}(\bar n)$ is used as a reminder that the distribution depends on both $A$ and $W$. 
Since the hafnian is independent of the parameters $w_k$, it is possible to express the derivative of the distribution in terms of GBS probabilities. Explicit calculations are done in Appendix \ref{a.haf} and the result is 
\begin{align}
\partial_{w_k} P_{A,W}(\bar n) 
&= \frac{n_k -\langle{n_k\rangle}}{w_k}  P_{A,W}(\bar n),
\label{e:dpw}
\end{align}
where $\langle n_k\rangle$ is the average number of photons in mode $k$, which can be calculated directly from the covariance matrix $V$:
\begin{equation}
\langle n_k\rangle=\frac{V_{k,k}+V_{k+m, k+m}-1}{2}.
\label{e:nk}
\end{equation}

The above can be generalized with a reparametrization of the weights, namely 
$w_k=w_k(\theta)$, so by the chain rule 
\begin{align}
\partial_\theta P_{A,W}(\bar n)  &= 
\sum_{k=1}^m \left(n_k -\langle{n_k\rangle} \right) 
\;	P_{A,W}(\bar n) \partial_\theta \log w_k
~.
\label{e:dpl}
\end{align}

From Eq.~\eqref{e:dpw} it is also possible to calculate the gradient of cost functions
\begin{equation}\label{e:cost_gradient}
\partial_\theta C(\theta)=\mathbb{E}_{\bar n \sim P_{A,W}(\bar n)}\left[\sum_{k=1}^m 
H(\bar{n})\,\left(n_k -\langle{n_k\rangle} \right) \partial_\theta \log w_k \right].
\end{equation}
Therefore, gradients can be obtained by sampling directly from the distribution to estimate this expectation value.

\subsection{Computing gradients classically}
We now show that the gradient of the KL divergence is straightforward to compute 
with the WAW parametrization. Indeed since $\partial_\theta \log P = \frac{\partial_\theta P}{P}$, from Eq.~\eqref{e:dpl}
the gradient can be written as 
\begin{align}
	\partial_\theta C_{\rm data}(\theta) &= -\mathbb{E}_{\bar n \sim P_{\rm data}}\left[ 
	\sum_{k=1}^m \left(n_k -\langle{n_k\rangle} \right) \partial_\theta \log w_k \right] 
	\nonumber
\\ &= -
	\sum_{k=1}^m \left(\langle n_k \rangle_{\rm data}-\langle{n_k\rangle}_{\rm GBS} \right) \partial_\theta \log w_k,
	\label{e:cdata_gbsgrad}
\end{align}
where we introduce the notation $\langle n_k\rangle_{\rm GBS}$ to distinguish the average photon number of 
Eq.~\eqref{e:nk} from the expectation value $\langle n_k\rangle_{\rm data}$, defined as $\langle n_k\rangle_{\rm data}=\mathbb{E}_{\bar n \sim P_{\rm data}}[n_k]$, or alternatively as
\begin{equation}
\langle n_k\rangle_{\rm data}=\frac{1}{T}\sum_{t=1}^Tn_k^{(t)},
\end{equation}
when the data distribution is defined in terms of a given dataset $\{\bar{n}^{(1)}, \ldots, \bar{n}^{(T)}\}$. 
When using the 
reparametrization of Eq.~\eqref{e:reparam}, the gradient is given by
\begin{align}
	\partial_{\theta} C_{\rm data}(\theta) &= 
	\sum_{k=1}^m \left(\langle n_k \rangle_{\rm GBS}-\langle{n_k\rangle}_{\rm data} \right) 
	f^{(k)}.
	\label{e:cdata_gbsgrad_repar}
\end{align}

This expression can be further simplified by defining
\begin{equation}
F_{\text{data}} := \sum_{k=1}^m\langle n_k\rangle_{\text{data}}f^{(k)},
\end{equation}
which depends only on the data and the choice of vectors $f$. We then have
\begin{equation}\label{e:gradclassical}
\partial_{\theta} C_{\rm data}(\theta) = \sum_{k=1}^m \langle n_k\rangle_{\text{GBS}} f^{(k)} - F_{\text{data}}.
\end{equation}
Once $F_{\text{data}}$ has been calculated, only $m$ terms $\langle n_k\rangle_{\text{GBS}} f^{(k)}$ need to be computed to obtain the gradient. This can be done in $O(m)$ time on a classical computer by using Eq.~\eqref{e:nk}. This is true even if sampling from the trained distribution is classically intractable.

Finally, we note that the log-likelihood function
\begin{equation}\label{e:log-lik}
\mathcal{L}(\theta)=\sum_{t=1}^T\log P_{A,W}(\bar{n}^{(t)}),
\end{equation}
which is also often used in unsupervised learning~\cite{kulesza2011learning}, is related to the cost function of Eq.~\eqref{e:cdata} by the formula
\begin{align}
	C_{\rm data}(\theta) &= \frac1T\sum_{t=1}^T \log\frac{1/T}{P_{A,W}(\bar n^{(t)})}\nonumber\\
	 &= -\frac{\mathcal L(\theta)}T-\log T,
\end{align}
and therefore
\begin{equation}
\partial_\theta \mathcal L(\theta) = -T\, \partial_\theta C_{\rm data}(\theta),
\end{equation}
meaning that the gradient formula of Eq.~\eqref{e:gradclassical} can be used to perform training for either of these two cost functions.

\subsection{GBS with threshold detectors}
Threshold detectors do not resolve photon number; they ``click" whenever one or more photons are observed. Mathematically, the effect of this detection on the GBS distribution can be described by the bit string $\bar x=(x_1, x_2, \ldots, x_m)$, obtained from the output $\bar n$ by the mapping
\begin{equation}
	x_k(\bar n) = \begin{cases} 
		0 & {\rm if ~~} n_k=0, \\
		1 & {\rm if ~~} n_k>0. \\
	\end{cases}
	\label{e:isingmap}
\end{equation}

The GBS distribution with threshold detectors is given by 
\begin{equation}\label{e:tor}
P_{A,W}(\bar x) = \frac{1}{\mathcal{Z}}\text{Tor}(X \mathcal{A}_W), 
\end{equation}
where $\mathcal{A}_W=A_W\oplus A_W$ and $\text{Tor}(\cdot)$ is the Torontonian function \cite{quesada2018gaussian}. This distribution does not factorize under the WAW parametrization as in Eq.~\eqref{e:wid}, which makes it challenging to compute exact gradients. Instead, we note that whenever 
$\langle n_k \rangle \ll 1$ it holds that
\begin{equation}
\langle n_k \rangle \approx \langle x_k \rangle_{\rm GBS},
\end{equation}
where we have implicitly defined $\langle x_k\rangle_{\rm GBS}$, the
probability of detecting at least one photon in mode $k$.
The latter can be computed efficiently as \cite{banchi2019molecular}
\begin{equation}
\langle x_k \rangle_{\rm GBS}
 = 1-\frac{1}{\sqrt{\det(Q^{(k)})}},
 \label{e:xtor}
\end{equation}
where $Q = (\id-X \mathcal{A})^{-1}$ and $Q^{(k)}$ is the submatrix obtained by keeping the $(k, k+m)$ rows and columns of $Q$. Under this
approximation, and assuming $\langle x_k\rangle\approx \langle n_k\rangle$, Eqs.~\eqref{e:cost_gradient} and \eqref{e:cdata_gbsgrad_repar}
can be updated to obtain
\begin{align}
	\partial_\theta C(\theta)&\approx\mathbb{E}_{\bar x\sim {\rm Tor}}\left[\sum_{k=1}^m 
H(\bar x)\,\partial_\theta \log w_k\left(x_k - \langle x_k\rangle_{\rm GBS} \right)  \right]\label{e:grad_tor},\\
	\partial_{\theta} C_{\rm data}(\theta)&\approx  \sum_{k=1}^m  \left[
 \langle x_k\rangle_{\rm GBS} -
 \langle x_k\rangle_{\rm data} 
\right]f^{(k)}\label{e:classic_tor},
\end{align}
where $\bar x\sim {\rm Tor}$ is a shorthand notation to say that $\bar x$ are sampled from Eq.~\eqref{e:tor},
and expectations
$ \langle x_k\rangle_{\rm data} $
are taken with respect to the data distribution. 
The opposite limit, $\langle n_k\rangle \gg1$ is studied in Appendix \ref{a:threshold}. A better approximation to the gradient in this limit is given by
\begin{align}
	\partial_\theta C(\theta) \approx
	\mathbb E_{ \bar x\sim {\rm Tor}}\left[ 
		H(\bar x) \;\sum_{k=1}^m v_k(\bar x) \partial_\theta \log w_k
		 \right],
	\label{e:grad_torlargen}
\end{align} 
where
\begin{align}
	v_k(\bar x) = 
		 {\max\left\{ \langle n_k\rangle (x_k-1), x_k - 
		 \langle n_k\rangle\right\} }. 
\end{align}
As we demonstrate in the Sec.~\ref{Sec:Apps}, these gradient formulas work sufficiently well in practice for training GBS distributions. These approximate formulas are also a biased estimator of the gradient, but it has been shown that convergence is 
expected even with some biased gradient estimators \cite{chen2018stochastic}.

\subsection{Quantum reparametrization} 
In this section we discuss an alternative training mechanism with a fixed 
Gaussian state. Before considering the application to GBS, we recall 
the general problem of stochastic optimization, namely to minimize 
the average value of a quantity that is estimated from sampled data. 
We assume that the data are distributed with a parametric probability
distribution $p_\theta(x)$ and the quantity to minimize is 
\begin{equation}
	C(\theta) = \mathbb{E}_{x\sim p_\theta(x)}[f(x,\theta)],
	\label{e:costC}
\end{equation}
where $f(x,\theta)$ is an arbitrary function that 
depends on the samples $x$ and possibly on the parameters $\theta$. 
The data distribution $p_\theta(x)$ changes if we update the 
parameters via training, so at each iteration a certain number 
of new samples must be obtained. 
Reparametrization is a common strategy \cite{kingma2013auto} to 
get an equivalent optimization problem to Eq.~\eqref{e:costC} with 
a $\theta$-independent distribution. It was recently employed 
to train generative models using quantum annealers \cite{vinci2019path}.
Reparametrization  is possible 
when a mapping $(x,\theta)\to z$ exists such that 
\begin{equation}
	p_\theta(x)dx = q(z) dz,
\end{equation}
with a new probability distribution $q(z)$. With the above definition 
we can write
\begin{equation}
	C(\theta) = \mathbb{E}_{z\sim q(z)} [ f(x(z,\theta),\theta)],
	\label{e:costZ}
\end{equation}
where data comes from a fixed, $\theta$-independent distribution. 
When the cost can be expressed this way, it is possible 
to get a fixed number of samples before training and optimize 
$C(\theta)$ without having to generate new samples after 
each iteration. Moreover, gradients obtained from Eq.~\eqref{e:costZ} 
typically have a lower variance. 

This strategy can be applied to the WAW parametrization because of the explicit form 
of Eq.~\eqref{e:GBSW}. More general parametrizations are studied in Appendix \ref{a:reptrick}.
Indeed, the cost function can be written in an 
alternative form where the weights are shifted away from the distribution as
\begin{align}
	C(\theta) &= \sum_{\bar n} H(\bar n) P_{A,W}(\bar n) \nonumber\\ &
= \sum_{\bar n} H_A(\bar n, W) P_A(\bar n),
\end{align}
where $P_A(\bar n)$ is just Eq.~\eqref{e:GBSW} with $W=\id$
and, from Eq.~\eqref{e:GBSW},
\begin{equation}
	H_A(\bar n, W) = H(\bar n) \sqrt{\frac{\det(\openone-A_W^2)}{\det(\openone-A^2)}} \prod_j \frac{w_j^{n_j}}{n_j!}~.
	\label{e:Hw}
\end{equation}
The extra numerical cost in computing $H_A(\bar n, W)$ is small, as determinants and powers 
can be efficiently computed numerically. 
Due to the formal analogy between the above equation and Eq.~\eqref{e:GBSW} we find 
\begin{equation}
	\frac{\partial H_A(\bar n,W)}{\partial w_k} =H_A(\bar n, W) \frac{n_k-\langle n_k \rangle}{w_k},
\end{equation}
and, analogously to Eq.~\eqref{e:cost_gradient},
\begin{equation}
	\partial_\theta C(\theta) = \mathbb{E}_{\bar n\sim P_A(\bar n)}\left[ \sum_{k=1}^m H_A(\bar n, W) 
	 \left(n_k -\langle n_k\rangle \right)  \partial_\theta \log w_k 
	 \right].
	 \label{e:costrepr}
\end{equation}
The advantage of the above is that we can always sample from the same reference state.
This approach may be used when there is a preferred choice for the $\mathcal A$ matrix, or when generating new samples is expensive. The next section 
discusses the opposite scenario.

\subsection{Projected subgradient method}\label{s:proj}

In the WAW reparametrization, the matrix $A$ is fixed and must be set at the beginning, while the diagonal weight matrix is updated. Here we discuss a more general strategy where $A$ is also updated at each iteration. 

When following the gradient, it is important that the resulting 
matrix $A$ always corresponds to a physical 
Gaussian state. As discussed before, a sufficient 
condition to enforce this constraint is to require that $0\leq w_k\leq 1$
for all $k$, which can be enforced via a convenient 
parametrization. An alternative is to use the 
projected subgradient method, commonly employed in constrained 
optimization problems \cite{boyd2003subgradient,banchi2019optimization}. For a generic 
parametrized matrix $A$, the update rule reads 
\begin{equation}
A \rightarrow \mathcal P[A-\eta \partial C],
\end{equation} 
where $\partial C$ is a matrix with elements $(\partial C)_{ij}=\partial_{A_{ij}} C$ and 
$\mathcal P[A]$ is a projection step that projects
$A$ to the closest matrix corresponding to a physical Gaussian state. 
The projection step is formalized explicitly in Appendix~\ref{a:project} as a semidefinite program. The complexity of performing this projection is comparable to matrix diagonalization.

We may now combine gradient rules in the WAW parametrization with the projected subgradient method and directly update the matrix $A$ 
during the optimization. As outlined in the following algorithm, the strategy
is to initialize weights to $w_k=1$, update them by gradient descent, then
project the new $WAW$ matrix to the closest physical state, leading to a new
matrix $A'$.

Formally, let $A^{(i)}$ be the matrix at step $i$.
From an initial choice $A^{(0)}$, each iteration performs the
following steps: 
\begin{enumerate}
\item Set $\theta$ such that $w_k(\theta)=1$ for all $k$, e.g., set $\theta_k=0$ for all $k$ when using $w_k(\theta)=\exp(-\theta^T f^{(k)})$. 
\item At step $i$ in the optimization, update the parameters $\theta$ using $\theta\rightarrow \theta - \eta\, \partial_{\theta} C(\theta) =: \theta_{\rm new}$, where $\partial_\theta C(\theta)$ is computed using the Gaussian state with 
matrix $WA^{(i)}W$. 
\item  Construct $A_W^{(i+1)}=W(\theta_{\rm new}) A^{(i)} W(\theta_{\rm new})$.
\item Set the updated matrix $A^{(i+1)}$ as
		\begin{equation}
			A^{(i+1)} = \mathcal P\left[A_W^{(i+1)}\right].
			\label{e:Aupt}
		\end{equation} 
\end{enumerate}
Since in general some of the weights $w_k$ in $W(\theta_{\rm new})$ will satisfy $w_k>1$ after updating the $\theta$ parameters, the matrix $A_W^{(i+1)}$ does not lead to a physical state, meaning the projection step is non-trivial 
and the entire $A$ matrix is updated during the optimization. 
As such, this algorithm may be used when there is no preferred choice for the matrix $A$, which can be learned through this procedure. 

\section{Applications \& Numerical experiments}\label{Sec:Apps}

Here we apply the results of previous sections to train GBS distributions. The first example is stochastic optimization, where the goal is to identify ground states of an Ising Hamiltonian. We show that gradient formulas and optimization strategies can be used to train the GBS distribution to preferentially sample low-energy states. In the second example, we consider an unsupervised learning scenario where data has been generated from a GBS distribution with a known matrix $A$ but unknown weights. We demonstrate in different cases that classical gradient formulas can be employed to train the GBS distribution to reproduce the statistics of the data. In all examples, sampling from the GBS distribution is performed using numerical simulators from The Walrus library \cite{gupt2019walrus}.

\subsection{Variational Ising Solver}
We study a classical Ising Hamiltonian 
\begin{equation}
H(\bar x) = -\sum_{i}h_i x_i -\sum_{ij}J_{ij}x_ix_j,
\end{equation}
where $\bar x=(x_1,x_2,\ldots,x_m)$ and $x_k=0,1$. Finding the ground state of $H(\bar x)$ is in general
NP-hard, and many known NP-hard models have a known Ising 
formulation \cite{lucas2014ising}. 
We are interested in finding a model distribution that samples the Ising ground state
with high probability. The output of GBS with threshold detectors is a vector $\bar x$ of binary variables, which is well suited for Ising problems, so we consider it here. The cost function for training is the average energy 
\begin{equation}
	E(W) = \sum_{\bar x} H(\bar x) P_{A,W}(\bar x)
	\equiv \mathbb{E}_{\bar x \sim P_{A,W}(\bar x)}\left[H(\bar x)\right], 
	\label{e:energy}
\end{equation}
where $P_{A,W}(\bar x)$ is the distribution of Eq.~\eqref{e:tor}. The gradient of this
cost function with respect to the weights $w$ can be approximated via 
Eq.~\eqref{e:grad_tor}, when $\langle n_k\rangle \ll 1$, and using Eq.~\eqref{e:grad_torlargen} 
when $\langle n_k\rangle \gg 1$. The exact gradient of $E(W)$, which requires photon-number-resolving detectors, is introduced in 
the Appendix \ref{a:isingnrd}, while the various approximations that lead to 
Eqs.~\eqref{e:grad_tor} and \eqref{e:grad_torlargen}  are discussed in Appendix~\ref{a:threshold}.

As a concrete example, we focus on the Ising formulation of the maximum clique problem. Given a graph $G=(V,E)$ with vertex set $V$ and edge set $E$, a clique is an induced subgraph such that all of its vertices are connected by an edge. The maximum clique problem consists of finding the clique with the largest number of vertices. The NP-complete decision problem of whether there is a clique of size $K$ in a graph can be rephrased as the minimization of the following Ising model
\cite{lucas2014ising}:
\begin{equation}
	H_{K}(\bar x) = c_V H_V(\bar x) + c_E H_E(\bar x),
	\label{e:ising}
\end{equation}
where $c_V,c_E$ are positive constants and 
\begin{align}
	H_{V}(\bar x) &= \left(K-\sum_{v\in V}x_v\right)^2, \\ 
	H_{E}(\bar x) &= 
	\frac{K(K-1)}2 - \sum_{(u,v)\in E} x_u x_v,
\end{align}
with binary variables $x_v=\{0,1\}$.
The above Hamiltonian has ground state energy $E=0$ if and only if there is a clique 
of size $K$; otherwise $E>0$. 
The corresponding NP-hard problem of actually finding the maximum clique can also be 
written as an Ising model, though the corresponding Hamiltonian $H$ is more complicated 
\cite{lucas2014ising}.

We show that the training of a GBS distribution,
with $A$ fixed as the graph's adjacency 
matrix, leads to a distribution that samples Ising ground states with high probability. 
The adjacency matrix provides a starting guess, while the weights 
are variationally updated to get closer to the actual solution. 

\begin{figure}[t]
	\centering
	\includegraphics[width=0.8\linewidth]{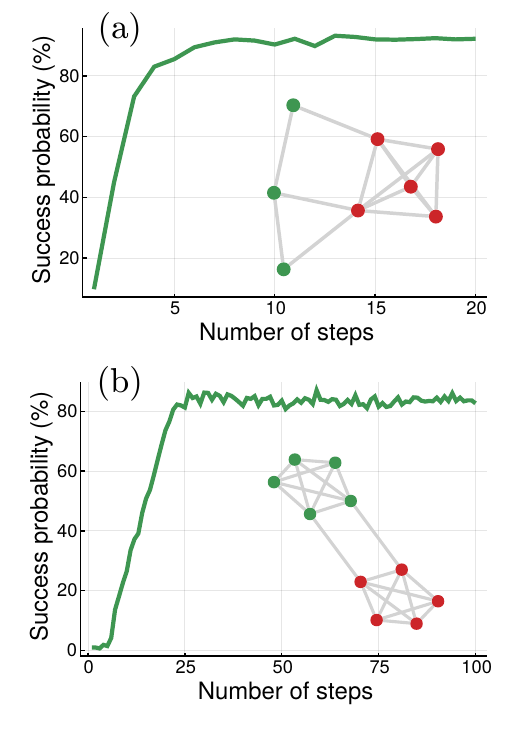}
	\caption{
		Success probability, namely the probability of sampling the bit string corresponding to the ground state of the 
		 Ising model \eqref{e:ising}, as a function of the number of steps, 
		for the displayed graph. The clique of size $K=5$ is shown in red. In
		(a) there is a single clique, while in (b) there are two degenerate cliques. 
		Training is done with $1000$ samples per iteration. 
	}%
	\label{fig:ising1}
\end{figure}

\begin{figure*}[t]
	\centering
	\includegraphics[width=0.95\linewidth]{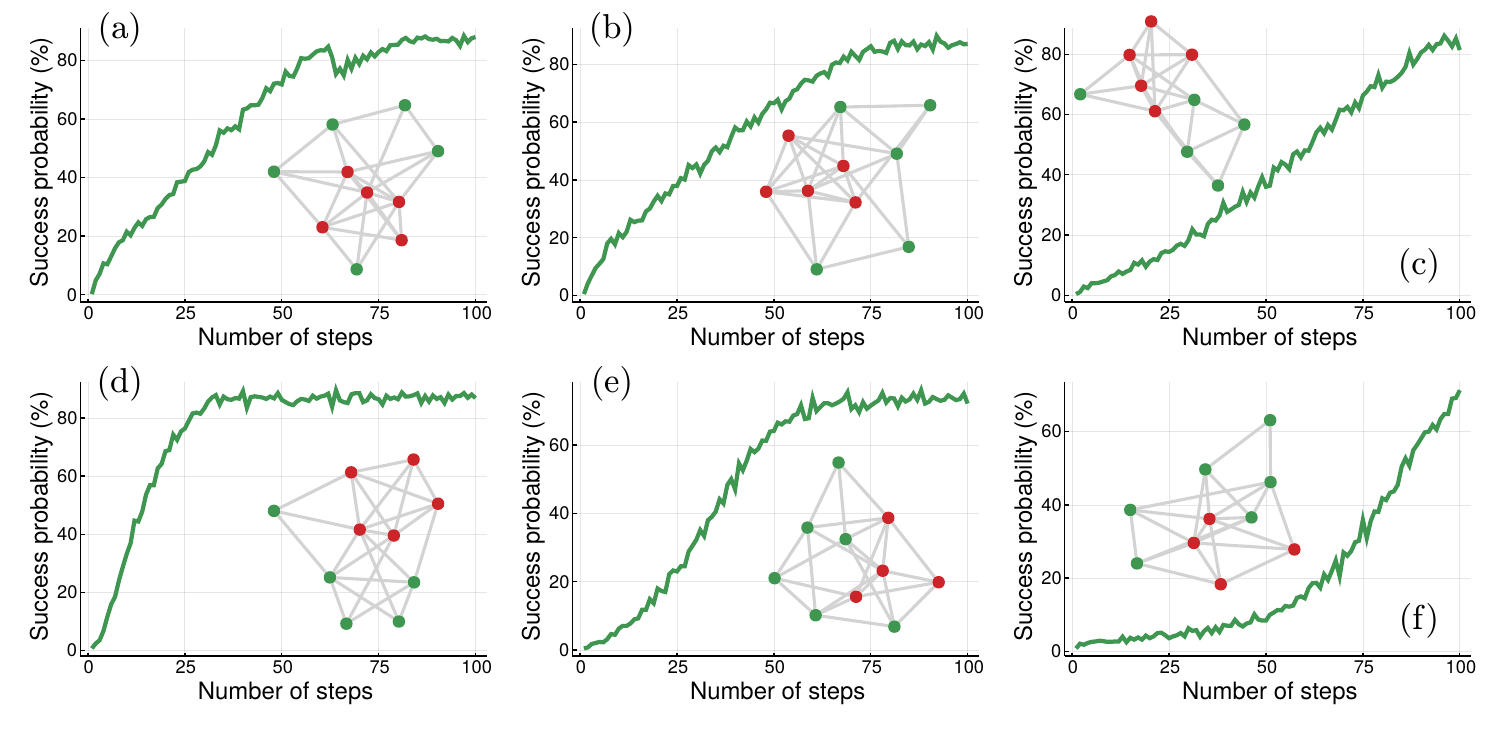}
	\caption{
		Success probability as a function of the number of steps, as in Fig.~\ref{fig:ising1},
		for the displayed graph. 
		Graphs (a),(b),(c) are random Barab\'asi-Albert 
		graphs with ten vertices, built starting from a clique of five vertices and attaching new vertices, each connected to three random nodes. 
		Graphs (d),(e),(f) are random Erd\H{o}s-R\`enyi 
		graphs with ten vertices and probability $p=0.5$ of adding an edge 
		between pairs of vertices. Clique sizes are either four or five. 
	}%
	\label{fig:ising2}
\end{figure*}

In Figs.~\ref{fig:ising1} and \ref{fig:ising2} we study the empirical success probability 
of sampling the bit string $\bar x_{\rm gs}$ that corresponds to the 
ground state of an Ising Hamiltonian with $c_V=2K$ and $c_E=1$.  The success probability is defined 
as the number of times that we get $\bar x_{\rm gs}$ in $1000$ samples, condition on observing $K$ clicks. To simplify the numerical calculations, the sampling
algorithm is configured to output a bit string with $\sum_k \langle x_k\rangle =K$, 
as explained below.
Training is done using an
estimation of the gradient as in Eq.~\eqref{e:grad_torlargen}, obtained with $1000$
samples per iteration.  
At each iteration, the physicality of the state is enforced by first mapping negative weights to zero,
then normalizing the weights so that they sum to one, 
and finally optimizing a coefficient $c$ in such a way that a Gaussian state with 
$A$-matrix $c(WAW)$ has $\sum_k \langle x_k\rangle =K$. Note that the weights are not reparametrized: they are directly optimized. The above operations take just a 
few milliseconds per operation, thanks to Eq.~\eqref{e:xtor}, and effectively 
implement a projection step as in Section~\ref{s:proj}.

In Fig.~\ref{fig:ising1}(a) we study a graph with eight vertices and a single clique of $K=5$ vertices. 
The probability of sampling the ground state 
of the Ising model is low, roughly 1.5\%, when sampling from an untrained distribution with $A$ equal to the adjacency matrix of the graph. However, using the WAW parametrization and 
updating the parameters via the momentum optimizer \cite{rumelhart1986learning}, we observe 
that the probability of sampling the ground state steadily increases 
and is above 85\% after a few iterations.

In Fig.~\ref{fig:ising1}(b) we study a more challenging example: a graph with ten vertices and two 
largest cliques of size $K=5$, for which the ground state of the corresponding Ising model is degenerate. Nonetheless, we observe that the training algorithm works almost as efficiently as 
with the simpler case of Fig.~\ref{fig:ising1}. During training, one of the two ground states is randomly selected and the algorithm keeps maximizing 
the sampling probability of that bit string without jumping to the other degenerate configuration. 
Runnning the algorithm multiple times we observe that upon convergence, both degenerate configurations can be obtained with essentially equal probability. 

In Fig.~\ref{fig:ising2} we switch to random graphs. The top row illustrates the effect of training for random 
Barab\'asi-Albert 
graphs, which are built starting from a clique of size $K=5$. These graphs are more complex than 
those of Fig.~\ref{fig:ising1} because they contain many cliques of size three and four. 
We observe that training allows jumping
from an initially low success probability to one higher than 80\% for sampling the ground-state configuration. 
The bottom row shows results obtained with random Erd\H{o}s-R\`enyi graphs with ten vertices, 
constructed by adding an edge with probability $p=0.5$. The graph in panel (d) has $K=5$, while 
the graphs in (e) and (f) have $K=4$. In all cases, the training procedure increases the probability of sampling the ground state configuration, from initial 
values close to 0\% to probabilities larger than 65\% after 100 iterations.

\subsection{Unsupervised learning}

In unsupervised learning, data is unlabelled and the goal is to train a model that can sample form a distribution induced by the data. Here, data is generated by sampling from a GBS simulator with threshold detectors that has been programmed according to a matrix $A_W=WAW$, where $A$ is the adjacency matrix of a graph, and a $W$ is a weight matrix. The data consists of one thousand samples from the distribution. For training, the weight matrix is assumed to be unknown, and the goal is to train a GBS distribution with the same $A$ to recover the weights that were used to generate the data.

\begin{figure*}[t!]
\includegraphics[width= 2\columnwidth]{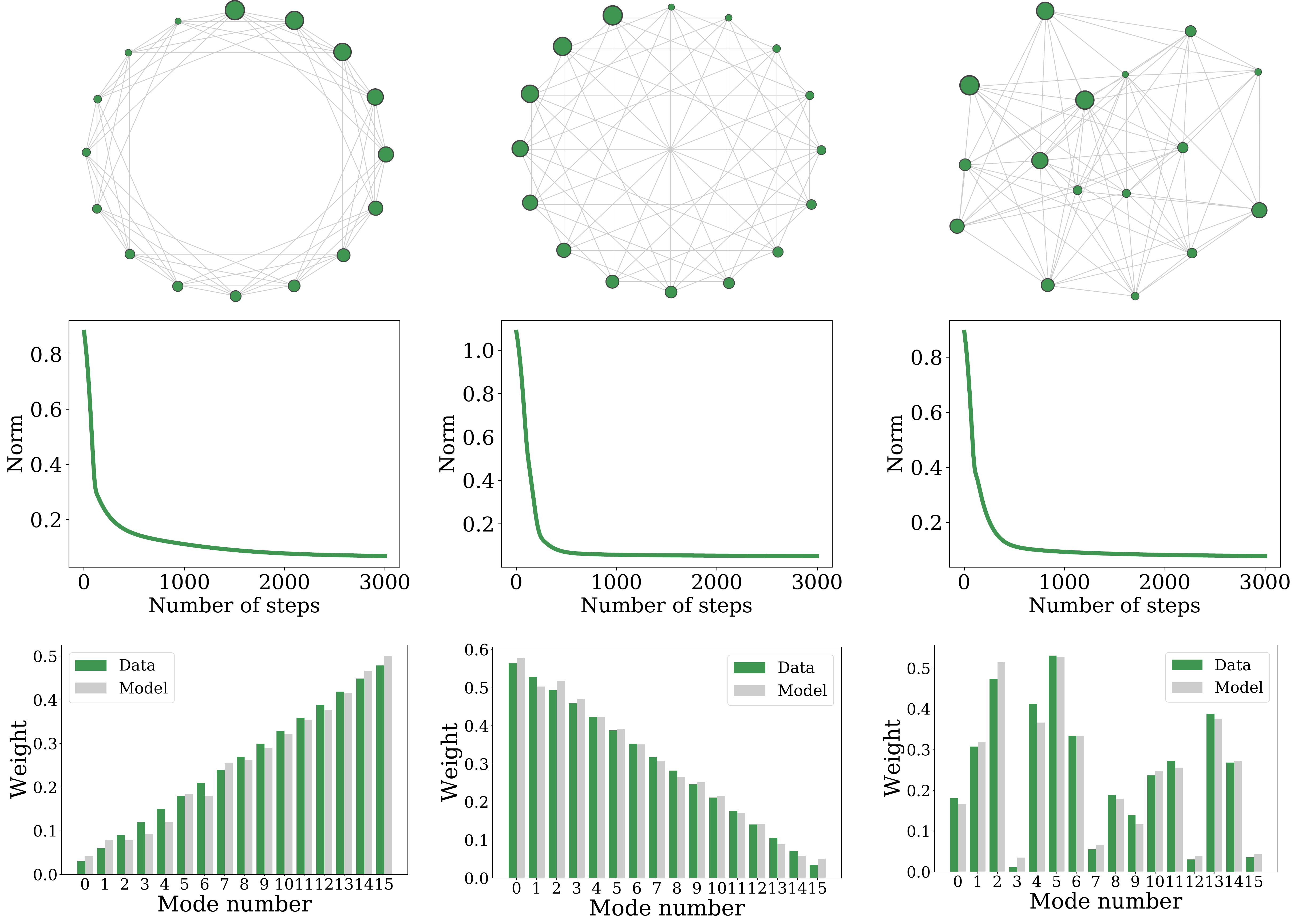}
\caption{Results of training a GBS distribution in an unsupervised learning scenario. (Top row): The graphs whose adjacency matrix $A$ is used to generate the training data from a GBS simulator. The first two graphs are circulant graphs, and the third is a random Erd\H{o}s-R\`enyi graph with edge probability $2/3$. The weights for the first graph are linearly increasing, they decrease linearly for the second graph, and for the random graph, they are chosen uniformly at random in the interval $[0,1)$. The size of the vertices is proportional to the weights of the $W$ matrix. The goal of training is to recover these weights.  (Middle row): The norm $\|W-W_{\text{model}}\|_2$ as a function of the number of steps in the optimization. Here $W$ is the weight matrix used to generate the data and $W_{\text{model}}$ is the weight matrix of the model. (Bottom row): Bar graph of the weights used to generate the data versus the weights of the trained model. }\label{fig:unsupervised}
\end{figure*}

We consider three examples. The first two cases explore circulant graphs, with linearly increasing and decreasing weights, respectively. These are configurations with a high degree of symmetry. The final example is a random Erd\H{o}s-R\`enyi graph with randomly-chosen weights, hence a less structured model. All graphs have sixteen nodes.

In each case, one thousand samples are generated as the training data, with a mean photon number $\langle n \rangle = 3$. For training, we employ the parametrization $w_k(\theta)=\exp(-\theta^T f^{(k)})$, where the vectors $f^{(k)}$ and parameter vectors $\theta$ are set to dimension $d=16$, equal to the number of vertices in the graph. The vectors are chosen to satisfy $f^{(k)}_l=\delta_{kl}$ such that $w_k(\theta)=\exp(-\theta_k)$. The cost function is the KL divergence, and we employ the approximate gradient formula of Eq.~\eqref{e:classic_tor}. We set a constant learning rate $\eta=0.1$ and find good results when initializing all weights to be small, so in all examples we set $\theta_k = 5$ for all $k$.

As shown in Fig.~\ref{fig:unsupervised}, optimization based on the gradient formula of Eq.~\eqref{e:classic_tor} works well for all examples. The weights of the model steadily and smoothly approach the data weights, until the weights at the end of training closely resemble those used to generate the training data. The entire training takes only a few seconds when running on a standard desktop computer. 

\section{Conclusions}\label{Sec:Conclusions}

We have derived a general formula for the gradient of the GBS distribution and have shown that, for 
specific parametrizations of the Gaussian state, the gradients of relevant cost functions take simple forms that can generally be efficiently estimated through 
sampling, or for specific situations, computed classically. 
Moreover, we have showcased this framework for training GBS distributions by applying it to problems in stochastic optimization and unsupervised machine learning. 

In stochastic optimization, we have introduced the variational Ising solver (VIS), 
a hybrid quantum-classical variational algorithm where the GBS device is used to 
generate samples that can be mapped to a set of binary variables. We have shown 
how to use the gradient formulas to train the GBS 
device in order to maximize the probability of sampling configurations that
correspond to the ground state of a classical Ising model. 
Many questions still remain open, especially in order to compare VIS with 
alternative algorithms, such as VQE or QAOA, for qubit-based computers. 
For instance, it would be interesting to study how to select the fixed $A$
matrix in the WAW parametrization, depending on the Ising Hamiltonian. 
Moreover, it remains to be proven if VIS can offer provable computational 
advantages against purely classical strategies, or whether any advantage is impossible.

In unsupervised learning, we have shown that for a specific parametrization,
the gradient of the Kullback-Leibler divergence between an unknown data distribution 
and the GBS distribution depends only the difference between 
the average photon numbers $\langle n_k\rangle$ of the two distributions. These averages can be computed classically, leading to fast training, which we show can be used to retrieve GBS parameters directly from data. To be the best of our knowledge, our results represent the first algorithms to 
variationally use near-term GBS devices to tackle optimization problems
in combinatorial optimization and machine learning.

\begin{acknowledgements}
	The authors thank N. Killoran and T. R. Bromley for valuable discussions and comments on the manuscript.
  L.B. acknowledges support by the program ``Rita Levi Montalcini'' for young researchers. 
\end{acknowledgements}

\appendix
\section{Gradient derivations}\label{a.haf}
We first focus on derivatives of Hafnians and show the following result:
\begin{prop}
	The derivative of $\partial_\theta {\rm Haf}(A(\theta))$ for a matrix $A$ that depends on a certain 
	parameter $\theta$ is given by 
\begin{align}
	\partial_\theta {\rm Haf}( A) 
= \frac12\sum_{j,ki}\sum_{i\neq j} (\partial_\theta  A)_{ij}  {\rm Haf}(
	A_{-j -i}),
	\label{e:gradprop}
\end{align}
where $A_{-j -i}$ is the submatrix of $A$ where rows $(i,j)$ and columns $(i,j)$ have been 
removed. 
\end{prop}
{\it Proof:} We
follow Ref.~\cite{kan2008moments}: given a set of non-negative integers $n_k$, where $N= \sum_{j=1}^m n_k$ is an even number, it holds that 
\begin{equation}
	{\rm Haf}( A_{\bar{n}})=\int \prod_{j=1}^m dx_j  \;\frac{e^{-\frac12 x^T  A^{-1} x} }{\det(2\pi A)^{1/2}} \;
	x_1^{n_1} \dots x_m^{n_m},
	\label{e:hafintegral}
\end{equation}
where $A$ is an $m\times m$ matrix, and $A_{\bar n}$ is constructed by repeating rows and columns of $A$ 
as discussed in Sec.~\ref{Sec:Gradients}.

Assume that the matrix $A=A(\theta)$ is parametrized by $\theta$. 
To calculate the derivative of the hafnian, we use Jacobi's formula 
\begin{equation}
	\partial_\theta \det( A) = \det( A) \Tr[  A^{-1} \partial_\theta  A],
	\label{e:jacobi}
\end{equation}
so from the chain rule
\begin{equation}
	\partial_\theta  \det( A)^{-1/2} = -\frac12 \det( A)^{-1/2} 
	\Tr[  A^{-1} \partial_\theta  A].
	\label{e:deta12}
\end{equation}
Moreover, 
\begin{align*}
\partial_\theta  e^{-\frac12 x^T  A^{-1} x} 
&= -\frac12 e^{-\frac12 x^T  A^{-1} x} (x^T\partial_\theta  A^{-1} x) \\
&=  \frac12 e^{-\frac12 x^T  A^{-1} x} (x^T A^{-1}\partial_\theta  A A^{-1} x) \\ 
&=  \frac12 \sum_{k,\ell} e^{-\frac12 x^T  A^{-1} x} x_k x_\ell ( A^{-1}\partial_\theta 
 A A^{-1})_{k\ell},
\end{align*}
where we used $\partial_\theta ( A^{-1})=- A^{-1}\partial_\theta  A A^{-1}$. 
Inserting the above equation in \eqref{e:hafintegral} we get 
\begin{align}
	\partial_\theta {\rm Haf}( A_{\bar{n}}) &= \frac12\sum_{k,\ell}
	\left( A^{-1}(\partial_\theta  A) A^{-1}\right)_{k\ell}  {\rm Haf}(
	 A_{\bar{n} + \bar e_k + \bar e_\ell})\nonumber\\
	 & - \frac12 
	\Tr[  A^{-1} \partial_\theta  A]\, {\rm Haf}( A_{\bar n}),
	\label{e:dhaf1}
\end{align}
where $\bar e_k$ is the vector with elements $(\bar e_k)_i = \delta_{ki}$. 
However, the above formula is not manifestly ``gauge'' invariant: since the hafnian does not depend on diagonal elements of the matrix, neither should its derivative. Below we show how the gauge symmetry can be explicitly restored. 
Without loss of generality, consider a matrix $A_{\bar{n}}$ with all $n_k=1$ that we simply 
call $A$. The extended matrix $A_{\bar e_k + \bar e_\ell}\equiv A_{\bar n +
\bar e_k + \bar e_\ell}$ in \eqref{e:dhaf1} takes the block form
\begin{equation}
	A_{\bar e_k + \bar e_\ell} = \begin{pmatrix}
		A_{11} & \dots & A_{1M} &\rvline & A_{1k} & A_{1\ell} \\
		\vdots & \ddots & \vdots &\rvline & \vdots & \vdots\\
		A_{M1} & \dots & A_{MM} &\rvline & A_{Mk} & A_{M\ell} \\
		\hline
		A_{k1} & \dots & A_{kM} & \rvline & A_{kk} & A_{k\ell} \\
		A_{\ell1} & \dots & A_{\ell M} & \rvline & A_{\ell k} & A_{\ell \ell} \\
	\end{pmatrix}~.
\end{equation}
Note that the above matrix has the elements $A_{kk}$ and $A_{\ell \ell}$ in 
off-diagonal positions, so they contribute to its Hafnian. 
Now we employ the Laplace-like expansion for the Hafnian 
\cite{barvinok2016approximating}
\begin{equation}
	{\rm Haf}(A) = \sum_{j\neq c} A_{jc} {\rm Haf}(A_{-j -c}),
	\label{e:lpexp}
\end{equation}
valid for any \emph{fixed} $c$, where $A_{-j-c}$ is matrix $A$ with rows $(j,c)$ and 
columns $(j,c)$ removed. 
Using  the expansion \eqref{e:lpexp} for 	
$ {\rm Haf}(A_{+\bar e_k + \bar e_\ell})$ when $c$ is the added column $\bar e_\ell$ 
(namely the $(M+2)$-th column) we get 
\begin{align}
	{\rm Haf}( A_{+\bar e_k + \bar e_\ell} ) &= 
		 A_{k\ell} {\rm Haf}( A) + 
		\sum_{j=1}^M  A_{j\ell} {\rm Haf}( A_{\bar e_k - j}),
\end{align}
where we used the fact that the index $j$ in \eqref{e:lpexp} takes $M+1$ values, as it 
runs from 1 to $M$ and to the copy of the $k$'s column. 
Inserting this equation into Eq.~\eqref{e:dhaf1} we get 
\begin{align}
	\partial_\theta {\rm Haf}(A) =& \frac12\sum_{k,j=1}^M 
	\left( (\partial_\theta  A) A^{-1}\right)_{jk}  {\rm Haf}(
	 A_{\bar e_k -j}) ~.
	 \label{e:tmpasdasd}
\end{align}
Using again Eq.~\eqref{e:lpexp} with $c$ equal to the added column $\bar e_k$ we get
\begin{align}
	{\rm Haf}( A_{+\bar e_k -j } ) &= 
		\sum_{i\neq j}  A_{ik} {\rm Haf}( A_{-i - j}),
\end{align}
Inserting the above in Eq.~\eqref{e:tmpasdasd} we get 
\begin{align}
	\partial_\theta {\rm Haf}( A) 
	 = \frac12\sum_{j,ki}\sum_{i\neq j} \left( (\partial_\theta  A) A^{-1}\right)_{jk}  A_{ik}{\rm Haf}(
	A_{- \bar e_j -\bar e_i}),
\end{align}
and the proposition follows. 
The above final form is independent of the diagonal elements of $A$, as desired. 
\hfill \qedsymbol

We now focus on the gradient of the GBS distribution in Eq.~\eqref{e:gradP}. Using \eqref{e:gradprop} with 
the matrix $A_{\bar{n}}$, we get  
\begin{equation}
	\partial_\theta {\rm Haf}( A_{\bar{n}}) = \frac12\sum_{i\neq j} 
	(\partial_\theta  A_{\bar n})_{ij} \; {\rm Haf}( A_{\bar{n} - \bar e_j -\bar e_i}) ~.
	 \label{e:dhaf}
\end{equation}
Finally to get $\partial_\theta \frac{1}{\mathcal Z} = \partial_\theta \sqrt{\det(\openone-X\mathcal A)}$ 
we can use \eqref{e:jacobi} to write
\begin{equation}
	\partial_\theta  \det(\mathcal B)^{1/2} = \frac12 \det(\mathcal B)^{1/2} 
	\Tr[  \mathcal B^{-1} \partial_\theta  \mathcal B].
	\label{e:deta123}
\end{equation}
Calling $\mathcal B = \id - X\mathcal A$, we we have
\begin{align}
	\Tr[\mathcal B^{-1}\partial_\theta \mathcal B] &= 
	-\Tr\left[\mathcal B^{-1}X\partial_\theta \mathcal A\right] \nonumber\\
	&=-\Tr\left[(\mathcal BX)^{-1}\partial_\theta \mathcal A\right]
	\cr &= -\Tr\left[\frac{1}{X - \mathcal A} \partial_\theta \mathcal A \right],
\end{align}
since $X=X^{-1}$. The above formula, together with \eqref{e:deta123} 
proves the resulting Eq.~\eqref{e:gradZ}.

For a pure state $\mathcal A = A\oplus A$ so we get 
\begin{equation}
	{P_A^{\rm pure}(\bar n)} = 
	\frac{ \sqrt{\det(\id - A^2)}}{\bar n! } {\rm Haf} (A_{\bar n})^2,
	\label{e.GBS0}
\end{equation}
and
\begin{align}
	\frac{\partial_\theta P_A^{\rm pure}(\bar n)}{P_A^{\rm pure}(\bar n)} =& -
	\frac12 
	\Tr\left[\frac{2A}{\id - A^2} \partial_\theta A \right] + 
	2 \frac{\partial_\theta{\rm Haf} (A_{\bar n})} {{\rm Haf} (A_{\bar n})}~.
	\label{e:dsqrtp}
\end{align}

Finally, we note that the formula \eqref{e:gradprop} for evaluating gradients of the Hafnian function 
allows us to compute also the gradient of matrix permanents. Indeed, 
from \cite{barvinok2016approximating} we have 
\begin{equation}
	{\rm per}(A) = {\rm Haf} \begin{pmatrix}
		0 & A \\ A^T & 0
	\end{pmatrix},
\end{equation}
so we can use Eqs.\eqref{e:gradprop} and \eqref{e:dhaf} to get the gradient of 
the matrix permanent.

\subsection{Gradients in the WAW parametrization}
Recall the GBS probability distribution in the WAW parametrization
\begin{align}
	P_{A,W}(\bar n) &= 
\sqrt{\det(\id - A_W^2)} \; {\rm Haf} (A_{\bar n})^2 
\prod_j  \frac{w^{n_j }_{j}}{n_j!}.
\end{align}

To write the gradient of the above distribution, we see that
\begin{equation}
	\frac{\partial_{w_k} 
	\prod_j  w^{n_j }_{j}}{\prod_j  w^{n_j }_{j}}
	 = \begin{cases}
		 \frac{n_k}{w_k}  & {\rm ~ ~ if ~ ~ } n_k>0, \\ 
		 0  &  {\rm ~ ~ otherwise}.
	 \end{cases}
\end{equation}
Then we get 
\begin{align}
	\partial_{w_k} P_{A,W}(\bar n)   &= \frac{n_k}{w_k}  P_{A,W}(\bar n) 
	-\\\nonumber&-\frac12 P_{A,W}(\bar n)  
	\Tr\left[\frac{2A_W}{\id - A_W^2} \partial_{w_k} W A W \right].
\end{align}
By explicit calculations 
\begin{align}
	\partial_{w_k} W A W  &= 
\frac12 w_k^{\frac12-1} (\ket k \bra k A W + W A \ket k\bra k )
\nonumber\\&=
\frac12 w_k^{-1} (\ket k \bra k W A W + W A W \ket k\bra k )
\nonumber\\&=
\frac12w_k^{-1} (\ket k \bra k A_W +A_W \ket k\bra k ),
\end{align}
we then obtain 
\begin{align}
	\partial_{w_k} P_{A,W}(\bar n) 
	&= \left(\frac{n_k}{w_k} -\frac{1}{w_k} 
	\bra k\left[\frac{A_W^2}{\id - A_W^2} \right]\ket{k} \right)
	{P_{A,W}(\bar n)}  \nonumber \\ 
	&= \frac{n_k -\langle{n_k\rangle}}{w_k} P_{A,W}(\bar n),
\label{e.dpw}
\end{align}
where $\langle n_k\rangle$ is the average number of photons in mode $k$.

\section{Weight updating}\label{a.weight}

\subsection{Spectral properties}
When $A$ has spectrum in $[-1,1]$ we show that, under some conditions, even the 
matrix $A_W$ has the same property. This corresponds to the requirement that
\begin{equation}
	|\bra{x}A_W\ket{x}| \le \bra xx\rangle ~ ~ {\rm ~~for~ each~} ~ \ket x~.
\end{equation}
Let $\ket y = W^{1/2}\ket x$ then 
\begin{equation}
	|\bra{x}A_W\ket{x}|  = |\bra y A \ket y \le \bra yy\rangle|  = |\bra x W\ket x|
	\le \bra x x\rangle,
\end{equation}
where we used the fact that the eigenvalues of $A$ are  smaller than one, while 
the last equality is true if 
\begin{equation}
	0\le w_k  \le 1~.
	\label{e:wcond}
\end{equation}
So if $A$ was a valid parametrization for a pure-state GBS distribution, then so is 
$A_W$, provided that the weights satisfy the above inequality. The conditions \eqref{e:wcond}
provide a sufficient condition for having a valid $A_W$ matrix, 
that in general is not necessary.

\subsection{Generalization to mixed states}\label{a:mixedvis}
A sensible generalization of the update rule in Eq.~\eqref{e:AW} is the following
\begin{align}\label{updateA}
\mathcal{A} \to \mathcal{A}_{\mathcal{W}} = \mathcal{W}^{1/2} \mathcal{A} \mathcal{W}^{1/2}.
\end{align}
where $\mathcal{W} = W \oplus W$. In the case where $\mathcal{A}$ is block diagonal then this rule indeed reduces to Eq.~\eqref{e:AW}, which is of course the desired limit behaviour.

Now we would like to argue that the transformation in Eq. \eqref{updateA} also maps a valid $\mathcal{A}$-matrix corresponding to a Gaussian state to another $\mathcal{A}_{\mathcal{W}}$ that corresponds to a Gaussian state. Recall that the covariance matrix $V$ of the Gaussian state is related to the $\mathcal{A}$-matrix as (recall Eq.~\eqref{e:amat})
\begin{align}
\mathcal{A} = X \left(\openone - \left[V+\openone/2 \right]^{-1} \right).
\end{align}
For $V$ to be a valid quantum covariance matrix it needs to satisfy the uncertainty relation
\begin{align}\label{e:uncertainty}
V + \frac{Z}{2} \geq 0,
\end{align}
where $Z = \sigma^z \otimes \openone_m$. 
The update equation for $\mathcal{A}$-matrices can be written in terms of the covariance matrix as
\begin{align}\label{updateV}
V &\to V_{\mathcal{W}},\\
 =&-\frac{\openone_{2m}}{2} + \left[\openone_{2m} - \mathcal{W}+\mathcal{W}^{1/2} \left(V+\frac{\openone_{2m}}{2}\right)^{-1} \mathcal{W}^{1/2} \right]^{-1}. \nonumber
\end{align}
One would like to show that the matrix $V_{\mathcal{W}}$ is a valid quantum covariance matrix if $V$ is a valid quantum covariance matrix, i.e. that it satisfies $V_{\mathcal{W}}+\tfrac{1}{2}Z \geq 0$. A simple way to show this is to first define the matrix $V^{\epsilon} = V+\epsilon \openone_{2m}$ which is always a valid quantum covariance matrix if $V$ is also in this set.
Then defining $V^\epsilon_\mathcal{W}$ to be the matrix obtained by letting $V \to V^{\epsilon}$ in Eq.~\eqref{updateV} one can easily show the following inequality
\begin{align}
&V^{\epsilon}_\mathcal{W} +\frac{\openone_{2m}}{2} \geq \\
&\begin{bmatrix}
\left( \openone_m -W + \epsilon^{-1} W  \right)^{-1} & 0 \\
0 & \left(\openone_m - W +(1+\epsilon)^{-1} W\right)^{-1}
\end{bmatrix}. \nonumber
\end{align}
assuming Eq.~\eqref{e:uncertainty} holds.
In the limit $\epsilon \to 0$, one has $V^\epsilon \to V$, $V_\mathcal{W}^\epsilon \to V_\mathcal{W}$ and
\begin{align}
\begin{bmatrix}
\left( \openone_m -W + \epsilon^{-1} W  \right)^{-1} & 0 \\
0 & \left(\openone_m - W +(1+\epsilon)^{-1} W\right)^{-1}
\end{bmatrix} \nonumber \\
\to \frac{\openone_{2m}}{2} - \frac{Z}{2},
\end{align}
thus showing that indeed $V_\mathcal{W} + Z/2 \geq 0$ and $V_\mathcal{W}$ is a valid covariance matrix.

\section{Variational Ising Simulation with Threshold Detectors } \label{a:threshold}
Numerical simulation of GBS is very complicated even for small scale problems, 
as the range of possible integer values $n_k$ is possibly unbounded. 
Moreover, from the experimental point of view, GBS requires NRDs, which are more 
complex and less efficient than threshold detectors. GBS with threshold detectors 
was introduced in \cite{quesada2018gaussian} and it was proven that the resulting 
sampling is still $\sharp$P hard. The use of threshold detector formally results in the 
mapping \eqref{e:isingmap}, namely the $k$th detector ``clicks'' only when $n_k>0$. 
We write $x_k=1$ in that case, and $x_k=0$ otherwise. 
The outcome is then a collection of binary variables $\bar x$ which are related 
to the number distribution via \eqref{e:isingmap}. As threshold detectors output a 
binary variable, they are well suited for Ising model formulation. 
In Appendix~\ref{a:isingnrd} we show that, when using number-resolving detectors,
exact gradients of the average energy can be obtained via an extension 
of the Ising model $H(\bar x) = H(\bar n)$, where all numbers 
$n_k$ are mapped to $x_k=0$ if $n_k=0$ and $x_k=1$ if $n_k\geq 1$.  
When using threshold detectors, this extension not required, as the output of the detectors is the desired
binary variable $x_k$. 
However, we also need to consider the other $n$-dependent terms in Eq.~\eqref{e:GwNRD}. 

Let $B_{\bar x} = \{ \bar n: \bar x(\bar n) = \bar x\}$ 
be the set of all possible integer sequences that produce the same 
binary string $\bar x$ via Eq.~\eqref{e:isingmap}.  Clearly, for fixed $x$, the set $B_{\bar x}$ 
contains infinitely many sequences $\bar n$.  The probability 
\begin{equation}
	p_{{\rm Tor},W,A}(\bar x) = \sum_{\bar n \in B_{\bar x}} p_{A,W}(\bar n),
	\label{e:torp}
\end{equation}
is the GBS probability with threshold detectors.
On the other hand, with these definitions, 
the energy gradient can be decomposed as 
\begin{equation*}
	\frac{\partial E(w)}{\partial w_k} = \sum_{\bar x } H(\bar x) \sum_{\bar n\in B_{ \bar x}}
	\; \frac{n_k - \langle n_k\rangle}{w_k} \; p_{A,W}(\bar n)~.
\end{equation*}
The aim is to separate the second sum for using \eqref{e:torp}.
Indeed, we may write 
\begin{equation}
	\sum_{\bar n \in B_{\bar x}}  n_k p_{A,W}(\bar n) = \tilde{n}_k(\bar x) \, p_{{\rm Tor},A,W}(\bar x) ,
\end{equation}
where 
\begin{equation}
	\tilde{n}_k(\bar x) = \begin{cases}
		\sum_{n_k} n_k p_{A,W}(n_k|\bar x,x_k{=}1) & {~~~\rm if~~ } x_k = 1, \cr
		0  & {~~~\rm if~~ } x_k = 0,
	\end{cases}
\end{equation}
and $p_{A,W}(n_k|\bar x,x_k{=}1)$ is the conditional probability 
of having $n_k$ photons given that the $k$th detector clicked and 
that the other detectors produced the outcome $\bar x$. 
With these definitions we finally get 
\begin{align}
	\frac{\partial E(w)}{\partial w_k}
&= \mathbb E_{\bar x\sim {\rm Tor}} \left[ 
	H(\bar x) \; \frac{\tilde{n}_k(\bar x) - \langle n_k\rangle}{w_k}
\right],
	\label{e.gradsamp}
\end{align}
where $\bar x\sim{\rm Tor}$ is a shorthand notation to write that $\bar x$ is sampled 
from \eqref{e:torp}.
The above gradient is still exact, as no approximations have been made so far. 
The expectation value $\langle n_k\rangle$ is simple to get in a closed form from 
the Gaussian covariance matrix, whereas the quantity $\tilde{n}_k(\bar x)$ 
is hard to estimate. 
Nonetheless, we can use the fact that $n_k\geq 1$ when $x_k=1$ to write
$	\tilde{n}_k(\bar x) \geq x_k$. The above implies 
\begin{align}
	\frac{\partial E(w)}{\partial w_k} &\geq \sum_{\bar x } 
	H(\bar x) \; \frac{x_k - \langle n_k\rangle}{w_k} \; p_{{\rm Tor},A,W}(\bar x)
	\cr &= \mathbb E_{\bar x\sim {\rm Tor}} \left[ 
	H(\bar x) \; \frac{x_k - \langle n_k\rangle}{w_k} \right],
	\label{e:gradsamplb}
\end{align}
namely the exact gradient is lower-bounded by a quantity 
that can be estimated with via GBS with threshold detectors. An alternative 
estimation of the gradient is via the approximation 
$	\tilde{n}_k(\bar x) \approx \max\{\langle n_k\rangle,1\}x_k $, so 
\begin{align}
	\frac{\partial E(w)}{\partial w_k} & \approx 
	 \mathbb E_{\bar x\sim {\rm Tor}} \left[ 
		 H(\bar x) \;\frac{\max\left\{ \langle n_k\rangle (x_k-1), x_k - 
		 \langle n_k\rangle\right\} }{w_k} \right],
	\label{e:gradsampapp}
\end{align}
While Eq.~\eqref{e:gradsamplb} is always a lower bound to the exact gradient, 
Eq.~\eqref{e:gradsampapp} is just an approximation. However, we found that in 
numerical experiments it performs very well. 

For GBS with number resolving detectors, Eq.~\eqref{e:energygradNRD} provides an unbiased estimator of the 
gradient, so converge can be exactly proven for stochastic gradient descent algorithms. 
On the other hand, Eqs.~\eqref{e:gradsampapp} and \eqref{e:gradsamplb} represent 
a biased estimator. Nonetheless, it has been shown that convergence is 
expected even with some biased gradient estimators \cite{chen2018stochastic}.

\section{General considerations on the quantum reparametrization trick}\label{a:reptrick}
To study a general form of the quantum reparametrization trick for GBS,
we write the cost function \eqref{e:cost} as
\begin{equation}
	C(\theta) = \sum_{\bar n} H(\bar n) P_{\mathcal{A}(\theta)}(\bar n)~.
	\label{e:costgbs}
\end{equation}
where $\mathcal A(\theta)$ is the $\theta$-dependent $\mathcal A$-matrix of a 
Gaussian state and $\bar n$ is a vector of numbers, where $n_i$ is the number of 
detected photons in mode $i$. The above cost function can be written using 
quantum operators as 
\begin{equation}
	C(\theta) = \Tr[ H \rho(\theta)],
	\label{e:Cparam}
\end{equation}
where  $\rho(\theta)$ is a quantum state (in general, not necessarily Gaussian) and
\begin{equation}
	H = \sum_{\bar n} H(\bar n)\ket{\bar n}\bra{\bar n}.
\end{equation}
If we expand the trace in the Fock basis, then for a Gaussian state with 
$\mathcal A$-matrix $\mathcal A(\theta)$ we get \eqref{e:costgbs}. Now assume that
\begin{equation}
	\rho(\theta) = \mathcal R_\theta[\rho_0],
\end{equation}
where $\mathcal R_\theta$ is a quantum channel, namely a completely positive trace preserving linear map,
and $\rho_0$ is a reference state that does not depend on $\theta$. Using the dual channel 
$\mathcal R^*_\theta$ 
we find 
\begin{equation}
	C(\theta) = \Tr[ \mathcal R_\theta^*(H)\rho_0],
	\label{e:Crepr}
\end{equation}
and 
\begin{equation}
	\partial_\theta C(\theta) = \Tr[ \rho_0\,\partial_\theta \mathcal R_\theta^*(H)]~.
\end{equation}
In \eqref{e:Cparam} the observable is $\theta$-independent, but the state $\rho(\theta)$ 
changes at each iteration. On the other hand, in Eq.~\eqref{e:Crepr} the quantum 
state is always the same and the observable is changed. 

GBS can be used for estimating the gradient in at least two cases
\begin{enumerate} \renewcommand{\theenumi}{\Roman{enumi}}
	\item When $\mathcal R_\theta$ maps diagonal states (in the Fock basis) to diagonal states. In that case
		\begin{equation}
			\mathcal R_\theta^*(H) = \sum_{\bar n} H_{\mathcal R}(\bar n,\theta)\ket{\bar n}\bra{\bar n},
		\end{equation}
		for some $H_{\mathcal R}(\bar n|\theta)$ that depends on $\mathcal R$. Calling $\mathcal A_0$ the
		$\mathcal A$-matrix of $\rho_0$ we find 
		\begin{equation}
			C(\theta) = \sum_{\bar n} H_{\mathcal R}(\bar n,\theta) p(\theta|\mathcal A_0),
		\end{equation}
		and
		\begin{align}
			\partial_\theta C(\theta) &= 	\sum_{\bar n} \partial_\theta H_{\mathcal R}(\bar n,\theta)\, p(\bar n|\mathcal A_0) =
			\\ &
			= \mathbb{E}_{\bar n \sim p(\bar n|\mathcal A_0)}[\partial_\theta H_{\mathcal R}(\bar n,\theta)]~.
		\end{align}
		Therefore, we can always sample from a reference state $\rho_0$ 
		to get the gradient. 
	\item When $ \partial_\theta \mathcal R_\theta^*(H)$ can be put in a diagonal Fock basis by a symplectic transformation $S(\theta)$, possibly 
		dependent on $\theta$. Namely if 
		\begin{equation}
			\partial_\theta \mathcal R_\theta^*(H) = \sum_{\bar n} h'(\bar n,\theta) \, S(\theta) \ket{\bar n}\bra{\bar n} S(\theta)^\dagger,
		\end{equation}
		then
		\begin{align}
			\partial_\theta C(\theta) &= 	\sum_{\bar n} h'(\bar n,\theta)\, p(\bar n|\mathcal A_{S(\theta)}) =
			\\ &
			= \mathbb{E}_{\bar n \sim p(\bar n|\mathcal A_{S(\theta)})}[h'(\bar n,\theta)],
		\end{align}
		where $\mathcal A_{S(\theta)}$ is the $\mathcal A$-matrix of the state $S(\theta)^\dagger \rho_0 S(\theta)$.
		Therefore, for each $\theta$ we can run a $\theta$-dependent GBS to estimate the gradient. 
\end{enumerate}

\section{Projection to the closest Gaussian state}\label{a:project}
We discuss the case of a pure Gaussian state with $\mathcal A=A\oplus A$ 
and $A^*=A$. In that case, a physical state is defined by the requirement 
that $A=A^T$ and that its spectrum lies in [-1,1]. The latter condition can be 
enforced by requiring that $A\pm\openone$ are positive semidefinite operators, 
so the projection step $\mathcal P[X]$ can be computed via semidefinite programming 
as 
\begin{align}
	&{\rm minimize~} \|X-A\|, \\
	&{\rm such~that~} A=A^T,  A\pm\openone\geq 0,
\end{align}
for a suitable norm $\|\cdot\|$.
Using the projected subgradients we can then update the parameters 
via \eqref{e:SGD} and \eqref{e:cost_gradient}, and then finding the closest 
Gaussian state via the projection. 


\section{Variational Ising Simulation with Number Resolving Detectors}\label{a:isingnrd}
The main difference between the configuration space $\bar x$ of an Ising problem and the 
possible outputs $\bar n$ of GBS is that 
$\bar x$ is a vector of  binary variables while $\bar n$ is made of arbitrary 
positive integers. There are many ways of defining a binary variable out of an integer. 
Here, we focus on the mapping \eqref{e:isingmap}, as it is naturally implemented 
experimentally by threshold detectors. By reversing that mapping we may 
extending the Ising model to arbitrary integer sequences via 
$H(\bar n) = H(\bar x(\bar n))$. With these definitions, the goal is then 
to minimize the average energy 
\begin{equation}
	E(w) = \sum_{\bar n} H(\bar n) p_{A,W}(\bar n)
	\equiv \mathbb{E}_{\bar n \sim p_{A,W}(\bar n)}\left[H(\bar n)\right]~. 
	\label{e:energyNRD}
\end{equation}
The gradient of the above energy 
cost function easily follows from Eq.~\eqref{e:dpw} (extension to the more general
\eqref{e:dpl} is trivial), and we find 
\begin{align}
	\frac{\partial E(w)}{\partial w_k} &= 
	\mathbb{E}_{\bar n \sim p_{A,W}(\bar n)}[G_k(\bar n,w)], 
	\label{e:energygradNRD}
	\\
	G_k(\bar n,w) &=  H(\bar n) \;
	\frac{n_k - \langle n_k\rangle}{w_k} 
	~.
	\label{e:GwNRD}
\end{align}
Therefore, we can estimate the gradient by sampling from the GBS devices, without 
calculating classically-hard quantities like the Hafnians. Indeed, from many sampled 
integer strings $\bar n$ we can easily calculate $G_k(\bar n|w)$ and update 
the weights following the stochastic estimation of the gradient.

\bibliography{biblio}
\end{document}